\documentclass[conference]{IEEEtran}

\usepackage{amsmath}
\usepackage[tight]{subfigure}
\usepackage[footnotesize]{caption} 
\usepackage{cite}
\usepackage{eurosym}
\usepackage{amsfonts}
\usepackage{graphics}
\usepackage{epsfig}
\usepackage{psfrag}
\usepackage{pstricks}
\usepackage{array}
\usepackage{shortvrb}
\usepackage{epsf}
\usepackage{graphicx}
\usepackage{rotating}
\usepackage{float}
\usepackage{color}
\usepackage{cite}
\usepackage{soul}
\usepackage{multirow}
\usepackage{acronym}
\usepackage{url}
\usepackage{breakurl}
\usepackage[breaklinks, hidelinks]{hyperref}

\acrodef{ess}[\textsc{ess}]{Energy Storage System}
\acrodef{dae}[\textsc{dae}]{Differential Algebraic Equation}
\acrodef{vsc}[\textsc{vsc}]{Voltage Sourced Converter}
\acrodef{coi}[\textsc{coi}]{Centre of Inertia}
\acrodef{fdf}[\textsc{fdf}]{Frequency Divider Formula}
\acrodef{wecs}[\textsc{wecs}]{Wind Energy Conversion System}
\acrodef{spvg}[\textsc{spvg}]{Solar Photo-Voltaic Generation}
\acrodef{tcl}[\textsc{tcl}]{Thermostatically Controlled Load}
\acrodef{hvdc}[\textsc{hvdc}]{High-Voltage Direct Current}
\acrodef{pll}[\textsc{pll}]{Phase-Locked Loop}
\acrodef{pmu}[\textsc{pmu}]{Phasor Measurement Unit}
\acrodef{rtds}[\textsc{rtds}]{Real-Time Digital Simulator}
\acrodef{emt}[\textsc{emt}]{Electromagnetic Transients}
\acrodef{tg}[\textsc{tg}]{Turbine Governor}
\acrodef{avr}[\textsc{avr}]{Automatic Voltage Regulator}
\acrodef{agc}[\textsc{agc}]{Automatic Generation Control}
\acrodef{gps}[\textsc{gps}]{Global Positioning Satellite}

\newcommand{\Dt}{\frac{d}{dt}}

\DeclareGraphicsExtensions{.eps,.png,.pdf}

\def \R {{\rm I\kern -2.2pt R\hskip 1pt}}

\newcommand{\PreserveBackslash}[1]{\let\temp=\\#1\let\\=\temp}
\newcommand{\bfg}[1]{\boldsymbol{#1}}

\IEEEoverridecommandlockouts
\begin{document}

\title{Frequency Quality in Low-Inertia Power Systems}

\author{ \IEEEauthorblockN{Taulant K\"{e}r\c{c}i,\IEEEauthorrefmark{1}
    \IEEEmembership{IEEE~Member}, Manuel Hurtado,\IEEEauthorrefmark{1}
    Mariglen Gjergji,\IEEEauthorrefmark{2} Simon
    Tweed,\IEEEauthorrefmark{1} \\Eoin Kennedy,\IEEEauthorrefmark{1}
    and
    Federico~Milano,\IEEEauthorrefmark{3}~\IEEEmembership{IEEE~Fellow}}\vspace*{0.3cm}
  \IEEEauthorblockA{
    \begin{tabular}{cc}
      \begin{tabular}{@{}c@{}}
        \IEEEauthorrefmark{1}
        Transmission System Operator,\\
        EirGrid, plc\\
        Ireland\\
      \end{tabular} &
      \hspace{0.3cm}
      \begin{tabular}{@{}c@{}}
        \IEEEauthorrefmark{2}
        Transmission System Operator, \\ OST, sha \\Albania\\
      \end{tabular} 
      \hspace{0.3cm}
      \begin{tabular}{@{}c@{}}
        \IEEEauthorrefmark{3}
        School of Electrical and Electronic Engineering, \\ University College Dublin \\Ireland\\
      \end{tabular} 
    \end{tabular}
  }
  \thanks{T.~K\"{e}r\c{c}i, M.~Hurtado, S.~Tweed, E.~Kennedy are with Innovation \& Planning office, EirGrid plc, Ireland; M.~Gjergji is with SCADA/EMS office, OST sha, Albania; and F.~Milano is with School of Electrical \& Electronic Engineering, University College Dublin, Dublin 4, Ireland. E-mails: \{taulant.kerci, manuel.hurtado, simon.tweed, eoin.kennedy\}@eirgrid.com, mariglen.gjergji@ost.al, federico.milano@ucd.ie.}%
  \thanks{F.~Milano is partly supported by the Sustainable Energy Authority of Ireland (SEAI) under project FRESLIPS, Grant No.~RDD/00681.}
}

\IEEEoverridecommandlockouts

\maketitle
\IEEEpubidadjcol

\begin{abstract}
  This paper analyses the issue of frequency quality in low-inertia power systems. The analysis is based on a real-world large-scale low-inertia power system namely, the All-Island transmission system (AITS) of Ireland and Northern Ireland currently accommodating up to 75\% of non-synchronous generation.  The paper is motivated by a recent trend of some frequency quality parameters such as the standard frequency deviation and the slow frequency restoration.  The paper first discusses the frequency control services currently in place to ensure frequency quality in the AITS.  An analysis of the frequency quality parameters of the AITS is then presented based on actual data.  The paper also discusses, through an illustrative example, the effectiveness of automatic generation control as a potential approach to keep frequency within the operational range. 
\end{abstract}


% =====================================
\section{Introduction}
\label{sec:intro}
% =====================================

% =====================================
\subsection{Motivation}
\label{Motivation}
% =====================================

The displacement of conventional synchronous generators by converter-interfaced generation such as solar and wind energy leads to reduced levels of system inertia.  Large-scale low-inertia power grids face many challenges, including but not limited to frequency stability, voltage stability, and converter-driven stability  \cite{8450880}.  While the power system community including both academia and industry is working towards addressing the above challenges, an emerging critical issue that, so far, has received little to no attention is frequency quality.  The objectives of this paper are to fill this gap and to raise awareness in the community on the topic.  

%\note{It would be useful to say here why frequency quality is a ``critical'' issue. At the end of the day, EirGrid handles very well the grid even if the frequency standard deviation is high. Just a sentence to justify the relevance of the frequency quality will be enough.}

\begin{table}[t!]
  \centering
  \caption{Frequency quality parameters of the CE and IE/NI \cite{entsoe, eirgrid}.}
  \label{tab:param}
  \begin{tabular}{cccccc}
    \hline
    Parameter  & CE & IE/NI  \\
    \hline
    Standard frequency range  & $\pm$ 50 mHz & $\pm$ 200 mHz \\ 
    Maximum instantaneous frequency deviation & 800 mHz & 1000 mHz\\
    Maximum steady-state frequency deviation & 200 mHz & 500 mHz\\
    Time to recover frequency & not used & 1 minute\\
    Frequency recovery range & not used & $\pm$ 500 mHz\\
    Time to restore frequency & 15 minutes & 15 minutes\\
    Frequency restoration range & not used & $\pm$ 200 mHz\\
    Alert state trigger time & 5 minutes & 10 minutes\\
    Maximum number of minutes & \multirow{2}{*}{15,000} & \multirow{2}{*}{15,000} \\
    outside the standard frequency range \\
    \hline
  \end{tabular}
\end{table}

\begin{figure}[t!]
  \begin{center}
    \resizebox{0.945\linewidth}{!}{\includegraphics{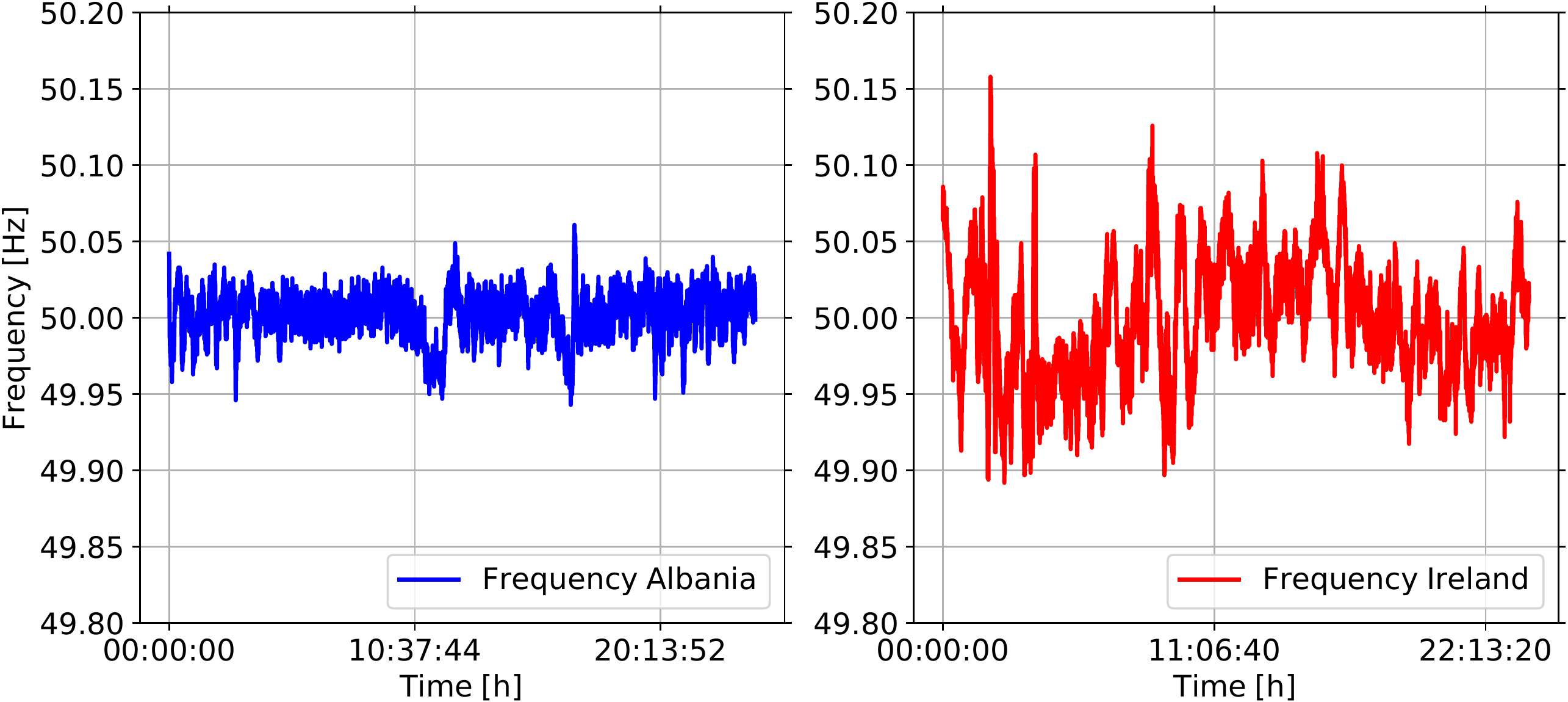}}
    \caption{Frequency in the CE and IE/NI power systems for 01.01.2021.}
    \label{fig:albania}
  \end{center}
  \vspace*{-0.3cm}
\end{figure}

% =====================================
\vspace{-2mm}
\subsection{Literature Review}
\label{sec:literature}
% =====================================

Transmission system operators (TSOs) in Europe (including EirGrid and SONI) define frequency quality in terms of different target/defining parameters.  Table~\ref{tab:param} shows the main parameters for the Continental European (CE) and Ireland/Northern Ireland (IE/NI) TSOs \cite{entsoe, eirgrid}.  
Keeping these parameters within limits help: (i) to better control the operation of the power system and prevent damage to plant and equipment; (ii)  keep the electric time on clocks that rely on counting the zero crossings; (iii) maintain the relevance of power system analysis that is generally performed at the nominal frequency; (iv) prevent motors from stalling; and (v) increase the trust of TSO customers and market participants on supply reliability and quality, etc. 
Note the wider range of most of the parameters of IE/NI compared to the parameters defined for the CE control area. For example, the standard frequency range for the CE and IE/NI control areas are $\pm$ 50 mHz and $\pm$ 200 mHz, respectively.  Such parameters make sense considering that the CE synchronous area accounts for around 435 GW of peak demand (largest synchronous electrical grid in the world)  compared to 6.9 GW of the All-Island transmission system (AITS) (i.e., it is harder to control frequency in a small power system).  To better illustrate these differences, Fig.~\ref{fig:albania} shows the frequency trace of the Albanian (part of CE) and AITS (IE/NI) for 01.01.2021. Clearly, and as expected, frequency fluctuates much more in the AITS compared to the CE power system.

Recent research has demonstrated that there is an almost linear relationship between renewables penetration and frequency variations \cite{KAZMI2022119565, 8783475, KERCI2020105819, 7891044}. This suggest that there is a need to deploy more and faster reserve resources to deal with the ever increasing penetration of stochastic and intermittent renewable sources.  For example, due to increased intra-interval fluctuations and limited ramping from generators, the frequency deviations in a provincial power system in China increased from 0.019 to 0.032 Hz from 2014 to 2020 (i.e., 68.4\% increase) \cite{9631168}.  In the same vein, reference \cite{9245548} focuses on the issue of frequency quality for Southwest China considering the operation data from asynchronous operation tests and automatic generation control (AGC).  It is also worth mentioning that controlling the frequency in power systems with high shares of photo-voltaic (PV) is a challenging task due to PV power dropping much faster than, for example, wind power (e.g., 60\% of the installed power capacity per minute when cloud passes) \cite{ZHANG2019809}.   

A way to meet frequency quality standards in low-inertia systems is that renewable energies and emerging technologies such as battery energy storage systems (BESS) provide frequency support.  In this context, references \cite{validation, 8606157} study the participation of a 30 and 10 MW wind farms in AGC and show their potential by testing the behavior against field measurements and experimental results, respectively.  On the other hand, the potential of solar PV providing AGC services under different conditions (solar resource intensity) is shown in \cite{300mw} through a successful 300 MW power plant test.  BESS is shown to respond 
well to AGC commands/set-points and provide secondary frequency 
regulation in \cite{8935195}.
%\cite{8935195, bess}

% \vspace{-2mm}
% =====================================
\subsection{Contributions}
\label{Contributions}
% =====================================

The specific contributions of this paper are the following:
\begin{itemize}
\item An analysis of frequency quality based on a real-world low-inertia system namely the AITS. 
\item Show through the analysis that while some frequency quality parameters have improved over the last years, others such as the standard deviation of the frequency is increasing linearly.
\item Propose different solutions to address the recent deterioration in frequency quality in the AITS.
\item Study the effectiveness of AGC to smooth frequency variations due to wind and load variations and noise.
\end{itemize}

% =====================================
\subsection{Paper Organization}
% =====================================

The remainder of the paper is organized as follows.  Section
\ref{sec:background} briefly describes the frequency control employed in the AITS.  Section
\ref{sec:case} provides the results of the frequency quality analysis of the AITS.  
Section
\ref{sec:agc} discusses the effectiveness of AGC in reducing the standard deviation of the frequency through an illustrative example.  Finally, conclusions and future work directions are given in Section
\ref{sec:conclu}.

% =====================================
\section{Frequency Control in the All-Island Transmission System }
\label{sec:background}
% =====================================

\begin{figure}[t!]
  \begin{center}
    \resizebox{0.8\linewidth}{!}{\includegraphics{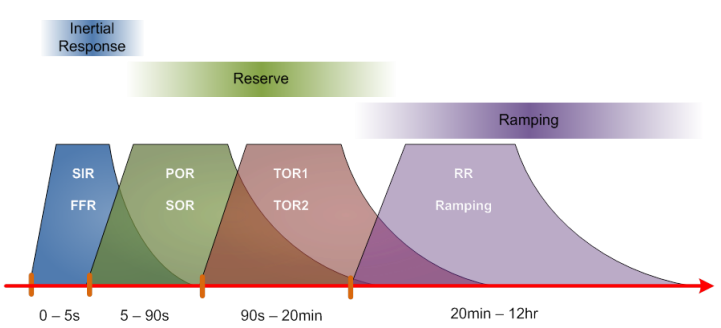}}
    \caption{Frequency control overview in the AITS \cite{control}.}
    \label{fig:control}
  \end{center}
  \vspace*{-0.3cm}
\end{figure}

Figure~\ref{fig:control} shows the current frequency control services employed in the AITS \cite{control}.  EirGrid and SONI have various frequency services to ensure frequency quality parameters remain within predefined limits. Such services include synchronous inertial response (SIR), fast frequency response (FFR) and primary, secondary and tertiary operating reserves (POR, SOR and TOR) as well as replacement reserves (RR) and ramping products.  It should be noted that there does not exist an automatic secondary frequency control (or AGC) in the AITS.  Instead, manual activations are currently the approach used in regulating system frequency proactively.  The vast majority of the contracted volumes of system services procured to date comes from conventional sources \cite{ds3}. However, as the AITS moves towards a reduced number of conventional units online by 2030, other technologies are expected to provide the majority of the services including BESS (approximately 650 MW installed and mostly used for system services rather than providing energy to the grid), demand response, wind and solar power. 

Active power control (APC) is another crucial frequency control service impacting frequency quality and currently in place in the AITS.  APC is a droop-based frequency control service and is mandatory for all dispatchable wind farms in Ireland \cite{code}.  It involves a selectable deadband setting of $\pm$ 200 mHz or $\pm$ 15 mHz. The default value of the deadband is $\pm$ 200 mHz but when the frequency control is challenging EirGrid changes remotely this deadband to $\pm$ 15 mHz.  In this mode, wind farms adjust their output much more dynamically and contribute to the control of system frequency under normal, pre-contingency conditions.  In the near future, it is expected that other technologies such as solar power will have APC functionality enabled in order to help maintain the frequency quality parameters.

% =====================================
\section{Frequency Quality in the All-Island Transmission System}
\label{sec:case}
% =====================================

The AITS is a synchronous island currently accommodating up to 75\% of non-synchronous generation at any point in time and is relaxing a number of operational constraints such as the minimum number of conventional generating units (from 8 to 7) and inertia (from 23 GWs to 20 GWs) to further increase the penetration of renewables \cite{policy}.  This transition involves dealing with different technical challenges such as ensuring power system stability and security. However, an emerging challenge is that of maintaining frequency quality parameters within acceptable limits.  The aim of this section is to analyse, through actual data, the frequency quality in the AITS.

% =====================================
\subsection{Frequency Deviations (Nadir/Zenith)}
\label{sec:parameters}
% =====================================

\begin{figure}[t!]
  \begin{center}
    \resizebox{0.8\linewidth}{!}{\includegraphics{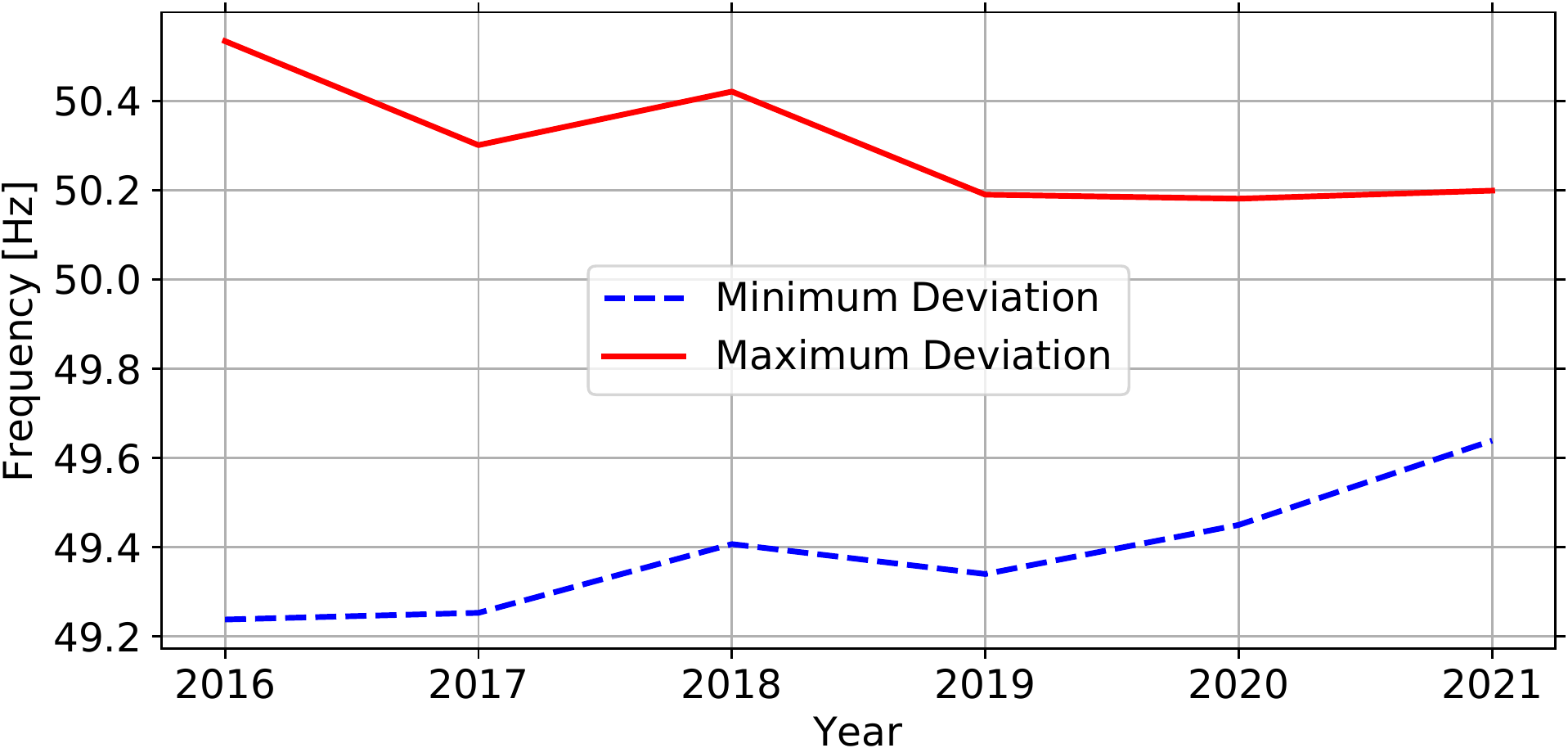}}
    \caption{Minimum and maximum frequency deviation evolution in the AITS.}
    \label{fig:minmax}
  \end{center}
  \vspace*{-0.3cm}
\end{figure}

This section focuses on the evolution of the maximum and minimum frequency deviations in the AITS over the last years.  Note that this limit is 1000 mHz according to Table ~\ref{tab:param}.  Figure~\ref{fig:minmax} shows the trend of these frequency parameters (i.e., frequency nadir and zenith).  It is interesting to notice that both parameters are within limits and improving in their performance.  The performance of these parameters is, in particular, strongly related to the reduced number of large generator trippings in the system and the response from wind, high voltage direct current (HVDC) interconnectors, demand response, and BESS.  

%

% =====================================
\subsection{Frequency Standard Deviation}
\label{sec:100mhz}
% =====================================

\begin{figure}[t!]
  \begin{center}
    \resizebox{0.85\linewidth}{!}{\includegraphics{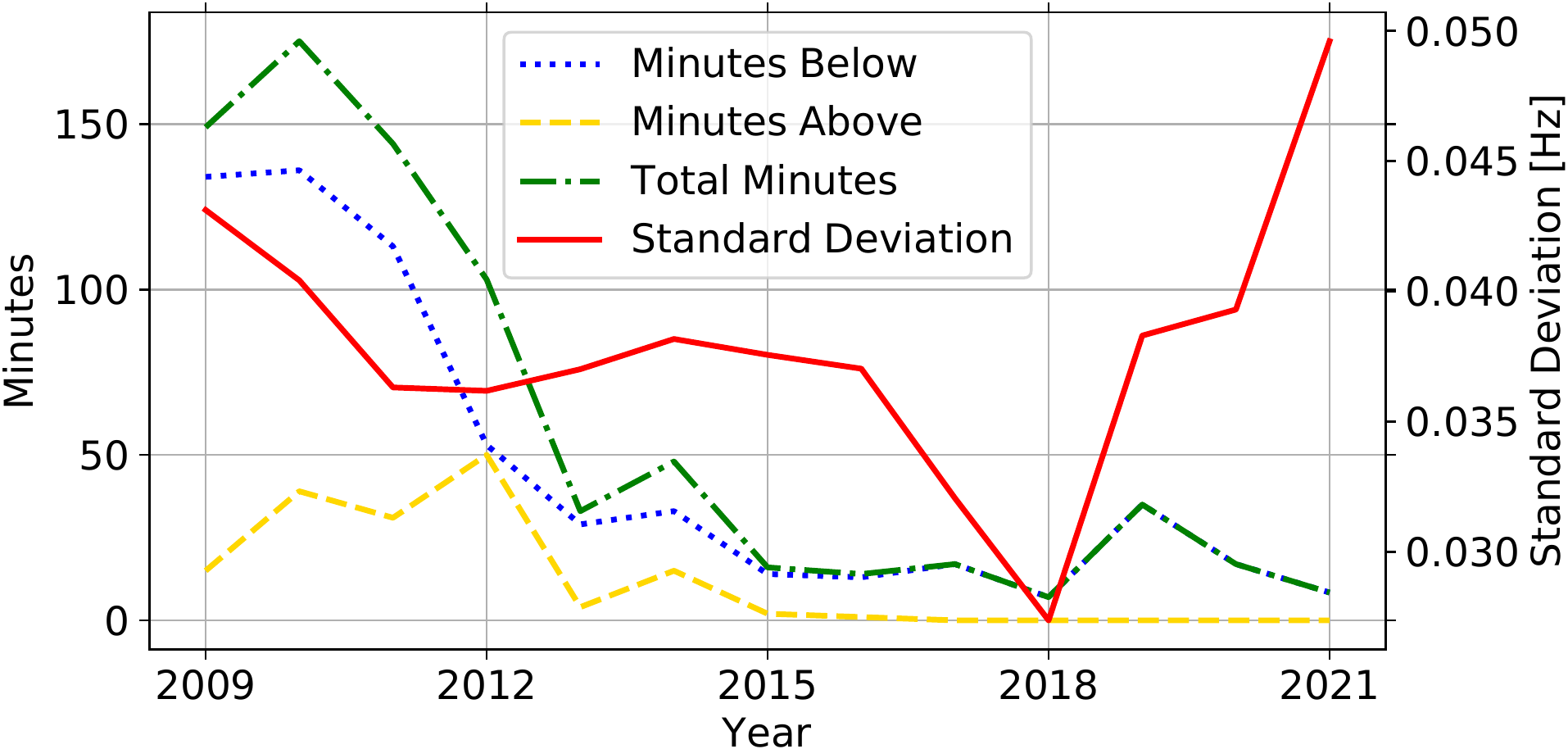}}
    \caption{Evolution of minutes outside the standard frequency range ($\pm$ 200 mHz) and standard deviation of the frequency over the last decade.}
    \label{fig:parameters}
  \end{center}
  \vspace*{-0.3cm}
\end{figure}

Figure~\ref{fig:parameters} shows the evolution of three main parameters namely minutes above and below the standard frequency range ($\pm$ 200 mHz) and the standard deviation of the frequency.  There are a number of factors that have led to a dramatic improvement in the quality of the parameters during 2009-2018, among others: (i) the change from verbal dispatch to electronic logged dispatch of generation units, leading to improved unit operator response; (ii) newer generating units in the latter years having the latest electronic governor controls; (iii) the retrofit of the electro-mechanical control systems on older generating units with modern electronic controls; (iv) the increase in system inertia as a result of more generating units to meet increases in system demand; and (v) the response of HVDC interconnectors and BESS (but also a reduction in the number of large generator trippings).
However, there has been an increase in the standard deviation of the frequency during 2019-2021.  This is mainly due to: (i) a reduction in regulating resources; (ii)  an increasing proportion of the reserves from inverter-based resources that are not configured to regulate frequency; and (iii) aging of conventional generating portfolio.  This trend could continue as more wind and solar power are integrated into the system and the operational policy evolves, i.e., reducing the number of conventional units online and, consequently, also reducing the inertia. 

\begin{figure}[t!]
  \begin{center}
    \resizebox{0.8\linewidth}{!}{\includegraphics{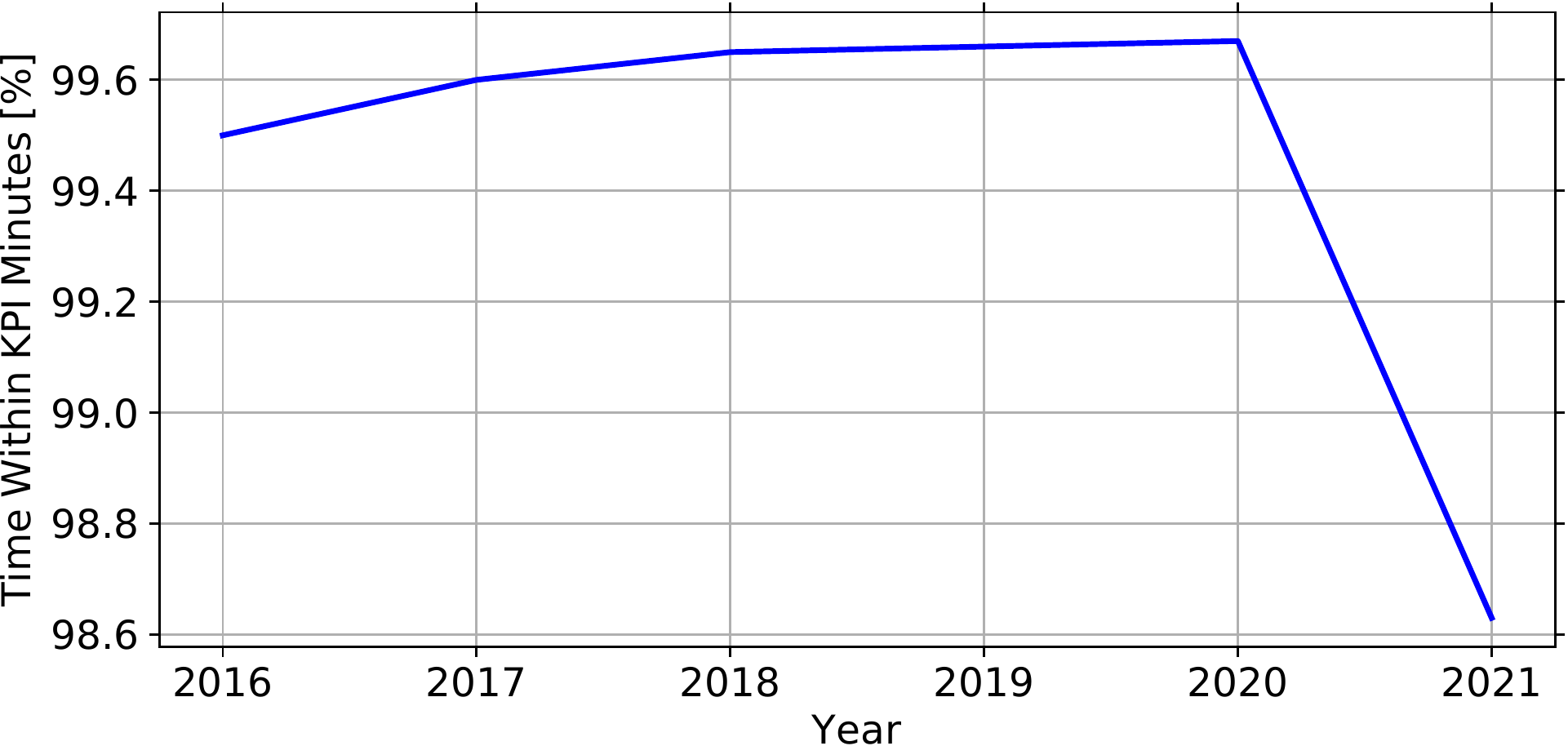}}
    \caption{Evolution of 100 mHz criteria over the last years in the AITS.}
    \label{fig:100mhz}
  \end{center}
  \vspace*{-0.3cm}
\end{figure}

While the EU Network Codes and the synchronous area operational agreement (Table \ref{tab:param}) require EirGrid to keep frequency within its standard range ($\pm$ 200 mHz), the national regulatory body in Ireland have put in place an incentive to keep frequency within an even tighter range namely $\pm$ 100 mHz for $\ge$ 98\% of the time \cite{cru}.  This is also known as the $\pm$ 100 mHz criteria.  Its evolution over the years is shown in Fig.~\ref{fig:100mhz}.  A deterioration can be seen for the last year.  Table \ref{tab:100mhz} shows the violation minutes for the last 12 months (i.e., greater during winter months mainly due to more wind).  This is a concern considering that, compared to 2020, there was less wind available in 2021.

\begin{table}[t!]
  \centering
  \caption{Frequency key performance indicator (KPI) statistics for 2021-2022 ($\pm$ 100 mHz criteria $\ge$ 98\% of time).}
  \label{tab:100mhz}
  \begin{tabular}{lccccc}
    \hline
    \multirow{2}{*}{Month} & \multirow{2}{*}{Minutes} & KPI \\
     & & Violation Minutes  \\
    \hline
    August 2021 & 44,640 & 372 \\
    September 2021 & 43,200 & 420  \\ 
    October 2021 & 44,640 & 862  \\ 
    November 2021 & 43,200 & 935  \\
    December 2021 & 44,640 & 1,803  \\
    January 2022 & 44,640 & 1,135  \\
    February 2022 & 40,320 & 1,066 \\
    March 2022 & 44,640 & 1,078 \\
    April 2022 & 43,200 & 411  \\
    May 2022 & 44,640 & 433  \\
    June 2022 & 43,200 & 333 \\
    July 2022 & 44,640 & 230 \\
    \hline
    Total minutes (12 months) & 525,600 & 9,078
\\
    \hline
    Time within KPI limits &  & 98.27\%
\\
    \hline
  \end{tabular}
\end{table}

There are a number of potential solutions to address the performance degradation of the frequency regulation in the last years (Figs.~\ref{fig:parameters} and \ref{fig:100mhz}), as follows: (i) increasing the minimum regulating/dynamic reserve requirement (which all currently comes from conventional generation with governor deadbands of $\pm$ 15 mHz) – this would seem a backward step given it will result in additional conventional unit commitment; (ii) narrowing frequency deadbands on BESS/HVDC interconnectors; (iii) updating rules for activation of wind farm APC (including NI wind farms when feasible) so that it is enabled more often, say when at minimum units – readily implementable and proven; (iv) implementing an AGC; and (v) introducing new market products to assist with frequency regulation.  In this paper, we explore the effectiveness of one of the options to improve frequency performance, that is, implementing an AGC (see Section~\ref{sec:agc} below).

% =====================================
\subsection{Frequency Recovery}
\label{sec:slow}
% =====================================

\begin{figure}[t!]
  \begin{center}
    \resizebox{0.8\linewidth}{!}{\includegraphics{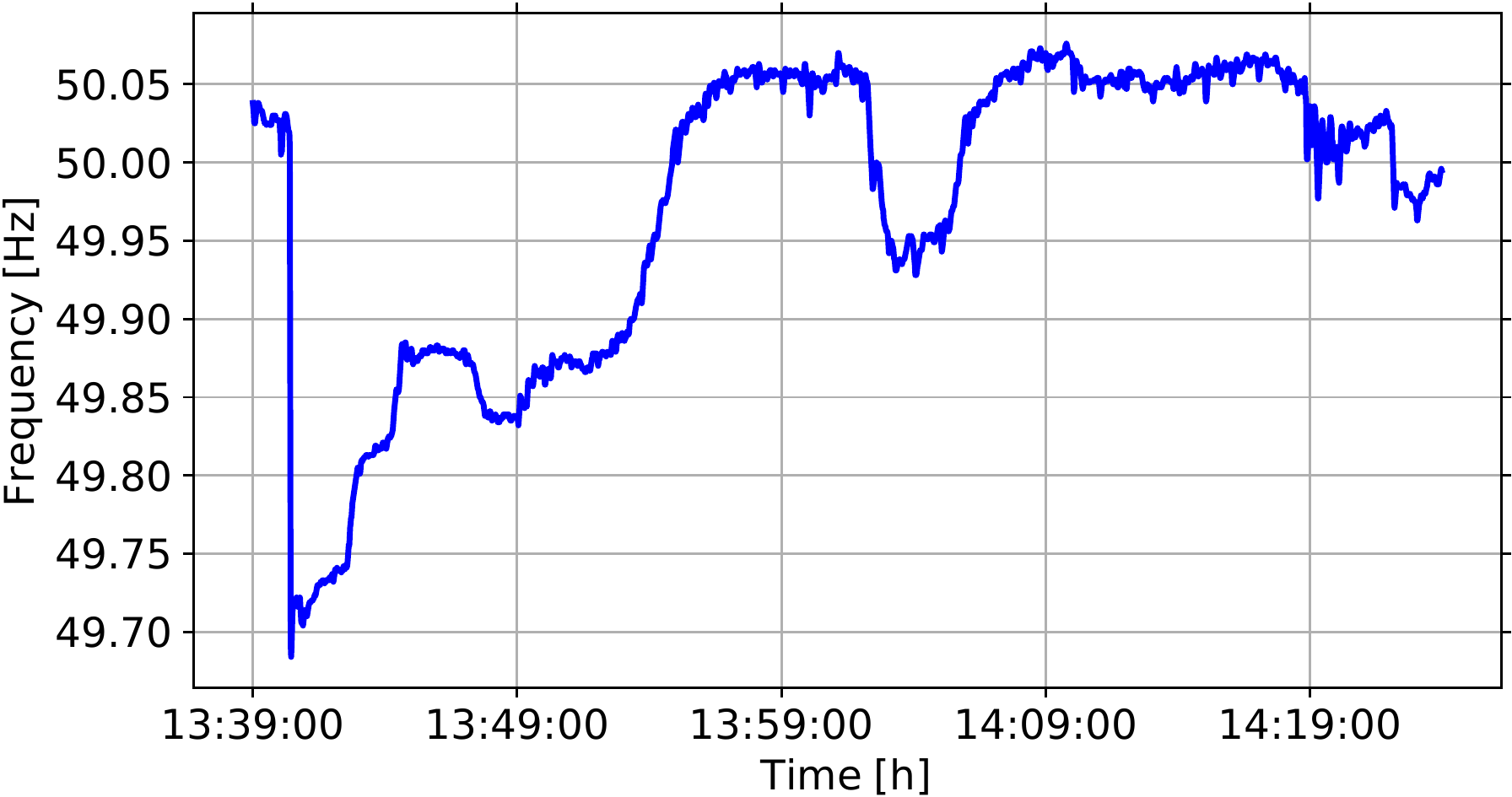}}
    \caption{LSI trip on August 2022.}
    \label{fig:ewic}
  \end{center}
  \vspace*{-0.3cm}
\end{figure}

This section illustrates the issue of the slow frequency recovery in the AITS.  With this aim, Fig.~\ref{fig:ewic} displays the frequency trace for a 2022 event (9th of August) where the largest single infeed (LSI) tripped from 530 MW import.  As can be seen, it took the frequency almost 15 minutes to recover to 50 Hz (i.e., time to restore frequency in Table \ref{tab:param}).  In particular, it is worth noticing that frequency recovers to around 49.87 Hz in almost 3 minutes but then stays there for a long time.  The fast recovery in the 3 minutes is mainly because of FFR from BESS (the majority of BESS installed in AITS have a trigger point of 49.8 Hz).   

% =====================================
\section{Illustrative example}
\label{sec:agc}
% =====================================

EU Network Codes and the national energy regulators require EirGrid and SONI to justify the need to install or not an AGC every few years \cite{entsoe}.  This section aims to illustrate the AGC performance in terms of long-term frequency quality enhancement using the IEEE 39-bus system.

% =====================================
\subsection{Stochastic Long-Term Power System Model}
\label{sec:stochastic}
% =====================================

Frequency quality is impacted by several dynamical processes starting from fast ones such as the inertial response of synchronous machines, the stochastic wind speed variations, to longer ones, such as primary and secondary frequency controllers of conventional generators.  To capture and model all of these dynamics, we consider a combined short- and long-term dynamic power system model represented by a set of hybrid non-linear stochastic differential-algebraic equations \cite{6547228}, as follows:
\begin{equation}
  \label{eq:hdae}
  \begin{aligned}
    \Dt{\bfg x} &= \bfg f( \bfg x, \bfg y, \bfg u, \bfg z, \Dt{\bfg \eta}) \, ,
    \\ \bfg 0 &= \bfg g(\bfg x, \bfg y, \bfg u, \bfg z, \bfg \eta) \, , \\
    \Dt{\bfg \eta} &= \bfg a( \bfg x, \bfg y, \bfg \eta) +
    \bfg b( \bfg x, \bfg y, \bfg \eta) \, \bfg \zeta \, ,
  \end{aligned}
\end{equation}
where $\bfg f$ and $\bfg g$ represent the differential and algebraic equations, respectively; $\bfg x$ and $\bfg y$ represent the state and algebraic variables, such as generator rotor speeds and bus voltage angles, respectively; $\bfg u$ represents the inputs, such as the schedules of synchronous generators; $\bfg z$ represents discrete variables; $\bfg \eta$ represents the stochastic characterization of wind speed as well as the volatility of load power consumption; $\bfg a$ and $\bfg b$ are the \textit{drift} and \textit{diffusion} of the stochastic differential equations, respectively; and $\bfg \zeta$ is the white noise.
To represent inertial and primary control dynamics, we consider conventional models of synchronous machines (4th-order models) and of their primary controllers, as well as dynamic models of wind power plants (5th-order doubly-fed induction generator) with inclusion of maximum power point tracker, voltage, pitch-angle, and frequency controls \cite{Milano:2010}.   

\begin{figure}[t!]
  \begin{center}
    \resizebox{0.8\linewidth}{!}{\includegraphics{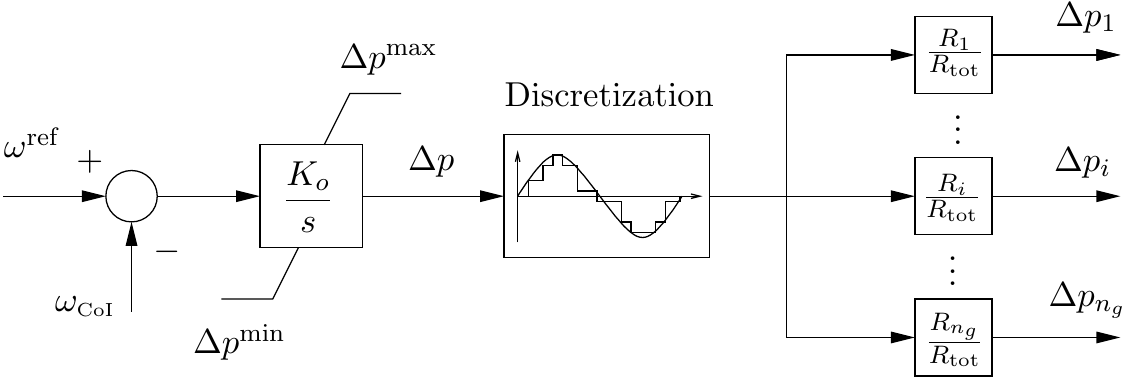}}
    \caption{Standard AGC.}
    \label{fig:agc}
  \end{center}
  \vspace*{-0.3cm}
\end{figure}

With regard to the long-term dynamics, the AGC is implemented as a centralized discrete controller in the control centers of TSOs and updates the power order set-points of dispatchable generators at certain time intervals, for example, every 4 seconds \cite{9361269}.  In this paper, we use the standard AGC scheme shown in Fig.~\ref{fig:agc}.  The AGC consists of an integrator with gain $K_o$ that aims to nullify the steady-state frequency error, in this case, the difference between the reference frequency $\omega^{\rm ref}$ and the measured frequency $\omega_{\rm CoI}$ (i.e., the center of inertia (CoI)), as follows: 
\begin{align}
  \label{eq:agc}
  \Dt{\Delta p} = K_{o}(\omega^{\rm ref}-\omega_{\rm CoI}) \, ,
\end{align}
where $\Delta p$ is the output of the integrator.  
To simulate the discrete nature of the AGC, $\Delta p$ is first discretized at given fixed-time intervals and then sent to each turbine governor (TG).  These signals ($\Delta p_i$) are proportional to the capacity of the machines and the TG droops ($R_i$) and normalized with respect to the total droop of the system:
\begin{align}
  \label{droop} R_{\rm tot} = \sum_{i=1}^{n_g}R_{i} \, .
\end{align}

% =====================================
\subsection{Simulation Results}
\label{sec:results}
% =====================================

The purpose of this section is to simulate the effectiveness of AGC to reduce frequency fluctuations.  The example is based on the IEEE 39-bus system and assumes a 25\% wind power penetration (i.e., replace three conventional generators with wind power plants).  Two scenarios are considered: (i) impact of stochastic noise (given by both load and wind); and (ii) scenario 1 plus the introduction of wind/load step and ramp variations \cite{KERCI2020105819}.  All the simulations in this section are performed using the software tool Dome developed by the last author \cite{6672387}.
\begin{figure}[t!]
  \begin{center}
    \resizebox{0.8\linewidth}{!}{\includegraphics{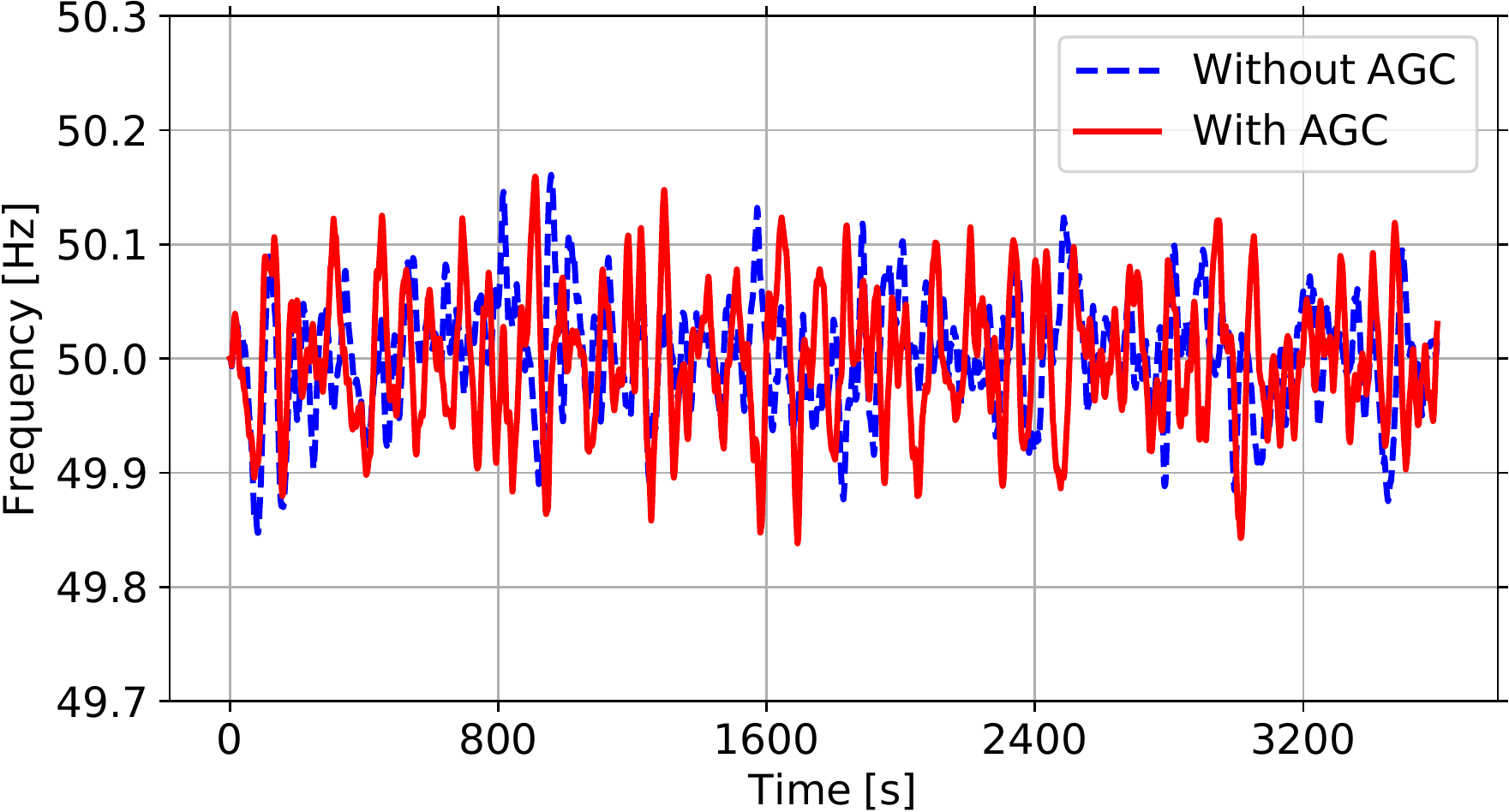}}
    \caption{Scenario 1: Impact of noise.}
    \label{fig:noise}
  \end{center}
  \vspace*{-0.3cm}
\end{figure}

\begin{figure}[t!]
  \begin{center}
    \resizebox{0.8\linewidth}{!}{\includegraphics{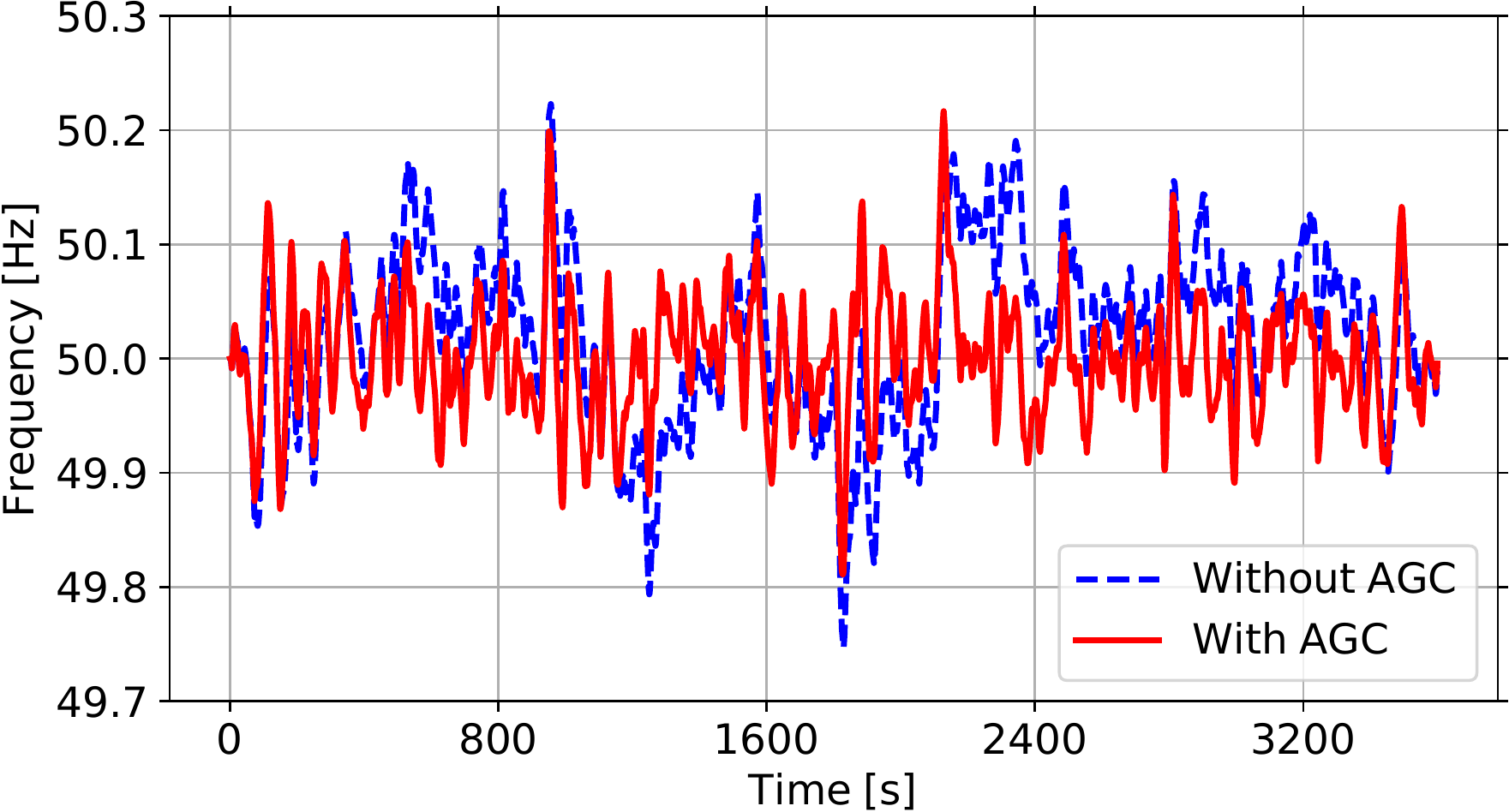}}
    \caption{Scenario 2: Impact of noise and wind/load step and ramp variations.}
    \label{fig:ramp}
  \end{center}
  \vspace*{-0.3cm}
\end{figure}

Figure~\ref{fig:noise} shows the results of the first scenario with and without the AGC.  In this scenario, the inclusion of the AGC does not appear to have any visible impact on frequency fluctuations. This is due to the fact that the AGC controller is slow compared to the dynamics of the noise (stochastic process).  On the other hand, Fig.~\ref{fig:ramp} compares the effect of AGC under both noise and wind/load step and ramp power variations.  Since wind/load ramp time scales are closer to that of the AGC, in this case, the inclusion of the AGC allows reducing frequency deviations.  Specifically, the standard deviations of the frequency with and without AGC are 0.05395 Hz and 0.0765 Hz, respectively.  These results indicate that an AGC implementation may be an option to improve frequency quality in the AITS in the future.  

% =====================================
\section{Conclusions}
\label{sec:conclu}
% =====================================

This paper discusses the issue of frequency quality in a real-world low-inertia power system, namely, the AITS.
The paper shows that while some frequency quality parameters have dramatically improved (e.g., minutes below and above $\pm$ 200 mHz) over the last decade, others have deteriorated.  In particular, the standard deviation of the frequency has increased linearly for the last three years.  The paper proposes different solutions to keep frequency within operational limits.  The potential effectiveness of one of the proposals, that is, installing AGC, is demonstrated through an example.  It is shown that AGC is an option to regulate frequency around the target value.

Future work will focus on testing the effectiveness of different AGC approaches on a model of the AITS.  This work will then feed in to the assessment of the range of solutions for managing frequency quality on the AITS.

% ======================================================================
%\bibliographystyle{IEEEtran}
%\bibliography{references}

% ======================================================================

\end{document}